\documentstyle[12pt]{article}
\def\beq{\begin{equation}}
\def\eeq{\end{equation}}

\begin{document}
  
\begin{titlepage}
\begin{flushright}
BA-99-55\\
September 3, 1999 \\
\end{flushright}

\begin{center}
{\Large\bf  R-Parity Violation:  \\
~~Origin of $\mu $-Term and Other Consequences
\footnote{Supported in part by  DOE under Grant No. DE-FG02-91ER40626
and by NATO, contract number CRG-970149.}
}
\end{center}   
\vspace{0.5cm}
\begin{center}
{\large Qaisar Shafi$^{a}$\footnote {E-mail address:
shafi@bartol.udel.edu} {}~and
{}~Zurab Tavartkiladze$^{b}$\footnote {E-mail address:
z\_tavart@osgf.ge} }
\vspace{0.5cm}

$^a${\em Bartol Research Institute, University of Delaware,
Newark, DE 19716, USA \\   

$^b$ Institute of Physics, Georgian Academy of Sciences,
380077 Tbilisi, Georgia}\\
\end{center}
  
\vspace{1.0cm}

\begin{abstract}

We propose a new mechanism in which the generation of the
supersymmetric
$\mu $-term as well charged lepton masses is closely tied to $R$-parity
violation
involving heavy vector-like families . A realistic example based
on $SU(3)_c\times SU(2)_L\times U(1)_Y$, supplemented by the symmetry
${\cal R}\times {\cal U}(1)$, is presented, where
${\cal R}$ (${\cal U}(1)$) denotes a continuous $R$ (flavor) symmetry. In
addition to the $\mu $-term,
the charged fermion mass hierarchies and mixings, as well as baryon number
conservation are also nicely explained. Bilinear $R$-parity violating
coupling
involving the first generation gives rise to neutrino mass relevant
for the small angle $\nu_e-\nu_s$ MSW oscillations, where $\nu_s $ denotes
a sterile
state. The atmospheric neutrino puzzle is resolved via maximal
mixing angle  $\nu_{\mu }-\nu_{\tau }$ oscillations.
The decay of the lightest neutralino (LSP) and leptogenesis
are briefly discussed.

\end{abstract}

\end{titlepage}

\section{Introduction}


The well known $\mu $ problem in supersymmetric (SUSY) theories is 
related to the so-called naturalness issue \cite{nc}, namely why 
(how) a given coupling (or mass) is tiny or zero, 
without any apparent symmetry reasons.
It is unclear how the magnitude of $\mu $ is in the $100$~GeV (or so)
range and not on the order of $M_{Planck }$ ($M_P$).
If by some discrete and/or continuous  symmetries
the $\mu $-term is forbidden at the  renormalizable level, its origin
can be explained either through higher order non-renormalizable 
operators \cite{nonren} or
non-minimal K\"ahler potential \cite{gm}. The latter case works 
within a framework in which SUSY breaking occurs through $N=1$
(supergravity)SUGRA,  and the contribution to the $\mu $-term 
is $\sim m_{3/2}$($\equiv $ gravitino mass
$\sim 10^3$~GeV), while other potentially large contributions \cite{kn}
are 
absent
from the theory with the help of suitable symmetries
\cite{chun}.
However, if SUSY breaking arises through gauge mediation, the 
induced $\mu $-term
through this mechanism will be heavily suppressed, since the gravitino
mass 
$\sim m_S \frac{M_M}{M_P}$ ($m_S\sim 10^3$~GeV is a SUSY scale and $M_M$
is a messenger mass $\stackrel{<}{_\sim } 10^{15}$~GeV in order to
solve supersymmetric flavor problem). In this case an alternative source for
the generation 
of the $\mu $-term with the desired magnitude is needed. 
Mechanisms for $\mu $-term generation within the gauge mediated SUSY
breaking scenarios were suggested in refs. \cite{gmed}.
A different possibility was recently discussed in \cite{choi} where
the $\mu $-term arises from  a new interaction at the TeV scale. 

In this paper we suggest a new mechanism for $\mu $-term generation whose
origin is related to`matter' (or R-parity) parity violation, and which
leads to some interesting
phenomenological  consequences \cite{rom}.
The proposed mechanism turns out to be quite general and can 
be used to build a variety of
realistic models. We begin our considerations by following the
naturalness criteria and assume that the $\mu $-term and Yukawa
couplings for the charged leptons all vanish at tree level. A crucial role
in
the generation of
both these couplings is played by the vector-like $SU(2)_L$ doublet states
$\overline E+E $.  In the limit when
$R$-parity is not violated the $\mu $-term is still zero. By 
violating
$R$-parity in the sector involving the heavy $\overline E+E$
superfields and integrating out the latter,
the charged lepton masses as well as the $\mu $-term can be
generated, which by suitable choice of the parity violating couplings
can have the desirable magnitude.

It turns out that the bilinear parity
violating operator(s) $h_ul$ can also be generated, so that the effective
low energy theory will have some
implications different from the minimal supersymmetric 
standard model (MSSM). For instance, the
lightest neutralino (LSP) becomes unstable and another 
candidate for cold dark matter
should be found. Also, one or more of the neutrinos can 
acquire tree-level mass which should be
properly suppressed (see later) in order to be phenomenologically
viable.

After presenting the mechanism we indicate the conditions
which should be satisfied and outline some clues which
can help realize it. We then turn to a specific example and 
consider a supersymmetric
standard model in which SUSY is broken through minimal $N=1$ SUGRA
and $R$-parity is replaced with the symmetry
${\cal R}\times {\cal U}(1)$, where ${\cal R}$ denotes a continuous
abelian
$R$-symmetry and ${\cal U}(1)$ is an anomalous flavor symmetry. 
The role of ${\cal R}\times {\cal U}(1)$ symmetry is three fold.
First, it forbids the (direct) $\mu $-term which is generated only
through the exchange of heavy states. Second, the ${\cal U}(1)$ symmetry
allows the possibility of naturally understanding the hierarchies of
fermion
masses and mixings. Finally, the ${\cal R}\times {\cal  U}(1)$  symmetry
also implies 
baryon number conservation, including higher dimensional operators. We 
note that in theories with $Z_2$
$R$-parity such as MSSM, the dimension five baryon number 
violating operators induce unacceptably
fast nucleon decay unless some mechanism for their suppression is employed
  
In our example $R$-parity 
(embedded in ${\cal R}\times {\cal U}(1))$ is violated only in the
sector of the first lepton
family. 
The smallness of tree level neutrino mass is also guaranteed by the   
${\cal R}\times {\cal U}(1)$ symmetry. For the generation of lepton masses we
introduce three pairs of $\overline E+E$ states, while the down quark Yukawa
couplings emerge through the exchange of three pairs of $\overline D^c+D^c$.
It is worth noting that the $\overline E$, 
$\overline D^c$ and $E, D^c$ states constitute
complete $\bar 5+5$ multiplets of the $SU(5)$, and because of this the
MSSM
unification of the three gauge couplings is retained in our model.
The resolution of the atmospheric and solar neutrino puzzles requires the
introduction of a  sterile neutrino state, which is kept light by
exploiting the
${\cal R}\times {\cal U}(1)$ symmetry \cite{st, d5ops}. The atmospheric
neutrino deficit is due to maximal $\nu_{\mu }-\nu_{\tau }$ 
mixing, while the solar
neutrino anomaly is resolved through the small angle $\nu_e-\nu_s$ 
MSW oscillations.

Nearly degenerate right handed neutrino states, which we invoke in the
neutrino sector, create a lepton
asymmetry through their decays, with the
CP asymmetry resonantly enhanced due to the mass degeneracy. 
This can explain the baryon asymmetry (which will be created from
lepton asymmetry during the electroweak phase transition) of our
Universe.

\section{Mechanism for $\mu $-Term Generation }
Let us consider the lepton sector and assume that the charged lepton masses
are generated through the exchange of some heavy states. 
For demonstration of the mechanism we will first consider the case of one
generation. The extension to all three generations will be straightforward.
We supplement the standard $e^c$, $l$ states with vector-like $\bar E+E$
pair, where $E$ has the same transformation properties as $l$, and
$\bar E$ is conjugate to $E$. Suppose that the direct coupling
$e^clh_d$ vanish by symmetry reasons, and consider the superpotential 

\beq
W=\lambda e^cEh_d+M\overline El+M_E\overline EE~,
\label{coup}
\eeq
where $\lambda $ is a dimensionless coupling, and $M$, $M_E$ are (heavy)
mass scales. Consider the mass matrix  

\begin{equation}
\begin{array}{cc}
 & {\begin{array}{cc}
\hspace{-5mm}~~l~ & \,\,~~~~E

\end{array}}\\ \vspace{2mm} 
\begin{array}{c}
e^c\\ \bar E
 \end{array}\!\!\!\!\! &{\left(\begin{array}{cc}
\,\,0  &\,\,\lambda h_d
\\ 
\,\,M~~   &\,\,M_E~~
\end{array}\right) }~.
\end{array}  \!\!  ~~~~~
\label{lep} 
\end{equation}
Assuming $M\stackrel{<}{_\sim }M_E$, the states $\overline E+E$ 
can be integrated out to yield a lepton mass

\beq
m_e\simeq \lambda \frac{M}{M_E}h_d~.
\label{lepmass}
\eeq
It is clear that (\ref{lepmass}) is valid below $M_E$.

The couplings in (\ref{coup}), (\ref{lep}) respect $R$-parity,
if $\overline E+E$ are treated as `matter' superfields.
For a moment let us assume that we do not have a priori $R$-parity in the
theory and include the following bilinear terms

\beq
W'=\tilde{m}\overline Eh_d+mEh_u~,
\label{bilin}
\eeq
where $m, \tilde{m}$ are some mass scales which, by assumption,
obey the relations:

\beq
m, \tilde{m}\stackrel{<}{_\sim }M_E~,~~~
M\stackrel{<}{_\sim }\tilde{m}~.
\label{rel}
\eeq
As we will see more precisely below, the first two relations
in (\ref{rel}) imply that the physical `light' electroweak Higgs doublets
mainly reside in $h_d$, $h_u$, while the last relation of $(\ref{rel})$
indicates that the `light' physical left-handed lepton doublet
state resides mainly in $l$. If the relations in $(\ref{rel})$ do not
hold, one
can redefine the appropriate superfields, so that the
conditions in (\ref{rel}) can be taken without loss of generality. 

Taking into account the couplings in (\ref{coup}), (\ref{bilin}),
consider the  mass matrix which would be relevant for the generation
of both the lepton mass and the $\mu $-term as well 
(we assume that the direct
$\mu $-term is also forbidden by symmetry reasons). Since the
states $h_d$ and $E$ mix, the latter
will develop a vacuum expectation value (VEV) which must be 
taken into account
during the analysis. The matrix takes the form

\begin{equation}
\begin{array}{ccc}
 & {\begin{array}{ccc}
\hspace{-5mm}~~~~l & \,\,~~h_d  & \,\,~~E

\end{array}}\\ \vspace{2mm}
\begin{array}{c}
e^c \\ h_u \\ \bar E
 \end{array}\!\!\!\!\! &{\left(\begin{array}{ccc}
\,\,0  &\,\, \lambda E &
\,\,\lambda h_d
\\
\,\,0   &\,\,0  &
\,\,m
 \\
\,\,M &\,\,\tilde{m} &\,\,M_E
\end{array}\right) }~.
\end{array}  \!\!  ~~~~~
\label{lep3}
\end{equation}
Taking account of (\ref{rel}), the states $\overline E+E$ can be integrated
out which generates the $\mu $-term
\beq
\mu \simeq \frac{m\tilde{m}}{M_E}~.
\label{mu}
\eeq
The scales $m, \tilde{m}, M_E$ should be chosen to obtain a
suitable value for $\mu $ (we will see later how symmetries can
help us achieve this). 
The lepton mass will be generated after
electroweak symmetry breaking and equals

\beq
m_e\simeq \frac{{\rm det }{\cal M}}{\mu M_E}=\lambda
\frac{M}{\tilde{m}}E~.
\label{lepmass1}
\eeq
From (\ref{lep3}) one can easily verify that
\beq
E\supset \frac{\tilde{m}}{M_E}h_d~.
\label{weight}
\eeq
Taking into account (\ref{weight}), from (\ref{lepmass1}) we recover
(\ref{lepmass}).

To summarize, by introducing $\overline E+E$ states in the theory and including
appropriate $R$-parity violating couplings, together with 
lepton masses one also can generate the $\mu $-term. The mechanism
suggested above is quite general and can be used to construct a variety of models.
Before constructing a realistic example, let us outline the conditions
which must be satisfied to avoid
conflict with phenomenology.

Note that together with the $\mu $-term, integration of $\overline E+E$
states also leads to the bilinear $R$-parity violating term

\beq
\mu_elh_u~,
\label{bilin1}
\eeq 
where, according to (\ref{lep3}), 

\beq
\mu_e\simeq \frac{Mm}{M_E}~.
\label{mu1}
\eeq
This term, in general, can create a non-zero sneutrino VEV,
which will lead to a neutrino
mass through mixings with neutralinos \cite{numass1}-\cite{numass4}.
Without any additional
mechanism of suppression, the neutrino mass is expected to be in the
$100$~GeV range! 

From (\ref{mu}), (\ref{mu1}) the generalized supersymmetric
`$\mu $-terms' have the form

\beq
W_{\mu }=(\mu_el+\mu h_d)h_u~,
\label{muterms}
\eeq
and one may think that, after suitable rotation of the doublets, only one
combination $h_d'$ will have coupling with $h_u$. However, this is not so,
because after SUSY breaking there also emerge the well known soft terms
$(m_s^2)_{ij}$ and $A, B$ ($i, j$ refer to superfields with identical transformation properties under
$SU(3)_c\times SU(2)_L\times U(1)_Y$). 
These soft terms , in general, are not universal and proportional (to $W$)
and therefore, in the minimization of the Higgs potential the
scalars of both $l$ and $h_d$ superfields participate.
As was shown in \cite{numass1, numass3}, the neutrino mass will vanish if 

\beq
(m_s^2)_{ij}=m_s^2\delta_{ij}~,~~~~~B_{\mu_i}\sim \mu_i~.
\label{univ}
\eeq
The neutrino mass is expected to be \cite{numass1}-\cite{numass4}

\beq
m_{\nu }\sim {\cal O}(100~{\rm GeV})\sin^2 \xi~,
\label{massnu}
\eeq
where the misalignment parameter

\beq
\sin \xi =\frac{B_{\mu_e}\langle h_d\rangle -B_{\mu }
\langle l\rangle }{(\langle h_d\rangle^2+\langle l\rangle^2)^{1/2}
(B_{\mu_e}^2+B_{\mu }^2)^{1/2}}~.
\label{misal}
\eeq
In the case of (\ref{univ}) one has the alignment 
\cite{numass1, numass3}

\beq
\frac{\mu_e }{\mu }=\frac{B_{\mu_e}}{B_{\mu }}=
\frac{\langle l\rangle }{\langle h_d\rangle }
\label{ali},
\eeq
so that $\sin \xi=0$ and the neutrino mass vanishes. 

An alternative way for suppressing the neutrino mass is to have a hierarchy
between $\mu_e$ and $\mu $, $B_{\mu_e }$ and $B_{\mu }$ 
\cite{numass1}-\cite{numass3}, \cite{hie2}

\beq
\mu_e \ll \mu~,~~~B_{\mu_e}\ll B_{\mu }~,~~~
\frac{\mu_e }{\mu }\sim \frac{B_{\mu_e }}{B_{\mu }}~. 
\label{hier}
\eeq
In this case $\sin \xi \sim \frac{\mu_e }{\mu }$, and the suppression 
factor
for the neutrino mass will be 
$\left( \frac{\mu_e }{\mu }\right)^2$.

The alignment in (\ref{univ}) can be achieved when SUSY is broken through
minimal $N=1$ SUGRA so that (\ref{ali}) is realized. 
For this case, with the mechanism
discussed above, the `$B_{\mu }$-terms' are generated after inclusion of the soft 
terms

\beq
V_{SB}=A_{\lambda }e^cEh_d+B_mh_uE+B_M\overline El+
B_{\tilde{m}}\overline Eh_d+B_{M_E}\overline EE~,
\label{soft}
\eeq
where the scalar components of appropriate superfields are assumed in
(\ref{soft}), and at $M_P$,

$$
A_{\lambda }=m_{3/2}A\lambda~,~~~~
B_m=m_{3/2}m~,~~~~B_M=m_{3/2}M~,~~~
$$
\beq
B_{\tilde{m}}=m_{3/2}\tilde{m}~,~~~~
B_{M_E}=m_{3/2}M_E~.
\label{soft1}
\eeq 
Taking into account (\ref{coup}), (\ref{bilin}), (\ref{soft})
it is easy to
verify that
after integration of $\overline E+E$ states, the $B_{\mu}$-terms are
generated

\beq
B_{\mu }=\frac{mB_{\tilde{m}}+\tilde{m}B_m}{M_E}~,~~~~
B_{\mu_e}=\frac{mB_M+MB_m}{M_E}~,
\label{bmu}
\eeq
and, using (\ref{soft1}), we see that $B_{\mu }$-terms
are aligned with the $\mu $-terms as in (\ref{ali}). 
However, both universality
and proportionality (\ref{soft1}) only hold at $M_P$. 
The alignment in MSSM is violated due to renormalization effects and 
one expects \cite{numass4}

\beq
\sin \xi ({\rm MSSM})\sim \frac{Y_b^2}{16\pi^2}\ln \frac{M_Z}{M_P}
\label{misb}~,
\eeq
where MSSM in parenthesis is inserted to remind the reader that
the estimate only holds if the field
content is identical with that of MSSM.

In our model there is a new source for misalignment.
Since 
$\mu $
and $B_{\mu }$-terms are generated at $M_E$,
the states $\overline E+E$ (and also the states $\overline D^c+D^c$, which
would generate down quark masses as in sect. (3.1) below) will provide
additional
contribution to the misalignment through 
renormalization between $M_P$ and $M_E$. 
Let us estimate this effect. The renormalization group (RG) equations
for the relevant parameters read

$$
16\pi^2\frac{dm_i}{dt}=m_i
\left (\lambda_{ji}\lambda_{ji}+3{\rm tr}(Y_uY_u^T)-3g^2-g'^2\right )
$$
$$
16\pi^2\frac{d\tilde{m}_i}{dt}=\tilde{m}_i
\left ({\rm tr}(\lambda \lambda^T)+
3{\rm tr}(\lambda_D\lambda_D^T)-3g^2-g'^2\right )
$$
$$
16\pi^2\frac{dM_i}{dt}=M_i
\left (-3g^2-g'^2\right )
$$
\beq 
16\pi^2\frac{dM_{E_i}}{dt}=M_{E_i}
\left (\lambda_{ji}\lambda_{ji}-3g^2-g'^2\right )~,
\label{rge}
\eeq
where $\lambda_D$ denotes couplings which will appear if the down quark masses
are also induced by integration of colored states circulating in
the loops.  
Taking into account (\ref{mu}) (\ref{mu1}), from (\ref{rge}) we obtain

\beq
16\pi^2\frac{d}{dt}\ln \frac{\mu_{e_i}}{\mu }=-{\rm tr}(\lambda \lambda^T)
-3{\rm tr}(\lambda_D\lambda_D^T)~.
\label{rgmis}
\eeq
Assuming that $\lambda ,\lambda_D \ll 1 $, from (\ref{rgmis}) one obtains

\beq
\left.\frac{\mu_{e}}{\mu }\right|_{M_P}-
\left.\frac{\mu_{e}}{\mu }\right|_{M_E}\simeq
\frac{1}{16\pi^2}[{\rm tr}(\lambda \lambda^T)+
3{\rm tr}(\lambda_D \lambda_D^T)]\ln \frac{M_E}{M_P}~.
\label{solrgmis}
\eeq

Clearly, analogous relations also hold for $B_{\mu }$-terms because
the $A_{\lambda }$ couplings are suppressed like $\lambda $ (see
(\ref{soft1})).
Therefore, the expected misalignment is estimated to be

\beq
\sin \xi \sim ~
\frac{1}{16\pi^2}[{\rm tr}(\lambda \lambda^T)+
3{\rm tr}(\lambda_D \lambda_D^T)]\ln \frac{M_E}{M_P}~.
\label{mis1}
\eeq
Note that in (\ref{mis1}), $\lambda , \lambda_D $ appear because they are 
the only
Yukawa couplings between $M_E$ and $M_P$ which can induce
misalignment. If they are small enough, the neutrino mass will
still be suppressed.

The second mechanism for suppressing the neutrino mass requires
a hierarchical structure (\ref{hier}), and can be realized through the 
flavor symmetries \cite{numass1}-\cite{numass3}, \cite{hie2}.
It is possible, of course, that these two suppression mechanisms of
tree level induced neutrino mass work together, in which case

\beq
\sin \xi \sim \frac{\mu_e}{\mu }
\frac{1}{16\pi^2}[{\rm tr}(\lambda \lambda^T)+
3{\rm tr}(\lambda_D \lambda_D^T)]\ln \frac{M_E}{M_P}~.
\label{mis2}
\eeq
In order to have neutrino mass in the range 
$\stackrel{<}{_\sim }0.1$~eV, one should have 
$\sin \xi \stackrel{<}{_\sim }3\cdot 10^{-7}$ (see (\ref{massnu})). 

Indeed, these two mechanisms are simultaneously present 
in the model presented below.

\section{The Model}

Consider the supersymmetric standard model with 
${\cal R}\times {\cal U}(1)$ symmetry and no $R$-parity a priori.
Under ${\cal R}$,  $W\to e^{{\rm i}R}W$,
$\phi_i\to e^{{\rm i}R_i}\phi_i$, where $R_i$ is the $R$-charge of the superfield
$\phi_i$. ${\cal U}(1)$ is a flavor symmetry which is anomalous.
As emphasized earlier, ${\cal R}\times {\cal U}(1)$ will be crucial for the
realization of the mechanism presented above, and for a natural explanation
of the hierarchies between fermion masses and mixings. The 
${\cal R}$ and  ${\cal U}(1)$ symmetry breaking scales also play a crucial role
in our considerations. 

Let us start with the description of ${\cal R}\times {\cal U}(1)$ symmetry
breaking. We introduce the singlet superfields $Z, \overline Z$, $X$ with
the
following transformation properties under ${\cal R}$ and ${\cal U}(1)$:

$$
{\cal R}~:~~~R_{W}=R~,~~~R_Z=\frac{2R}{5}~,~~~
R_{\overline Z}=-\frac{R}{5}~,~~~R_X=0~,
$$ 
\beq
{\cal U}(1)~:~~~Q_Z=q~,~~~~Q_{\overline Z}=-q~,~~~~Q_X\neq 0~.
\label{ch}
\eeq
The charges $q$, $Q_X$ are not fixed for the time being. However, let us
note that the neutrino sector helps fix these charges (see (\ref{cond}))  
in such a way that the single allowed term in the scalar superpotential 
involving
the $Z, \overline Z, X$ superfields, is

\beq
W_s=M_P^3\left(\frac{\overline ZZ}{M_P^2} \right)^5~.
\label{scsup}
\eeq 

In the unbroken SUSY limit the VEVs $\langle Z\rangle$,
$\langle \overline Z\rangle$ are zero. After SUSY breaking 
through minimal $N=1$ SUGRA, together with the soft terms

\beq
V_m=m_{3/2}^2(|Z|^2+|\overline Z|^2)~,
\label{softz}
\eeq
one finds:

\beq
\frac{|\langle Z\rangle|}{M_P}=
\frac{|\langle \overline Z\rangle |}{M_P}\equiv \epsilon_G\sim
\left(\frac{m_{3/2}}{M_P} \right)^{1/8}\simeq 10^{-2}~,
\label{vevz}
\eeq
where we have taken $m_{3/2}=10^3$~GeV, $M_P=2.4\cdot 10^{18}$~GeV.
The scale of ${\cal R}$ symmetry breaking is therefore close to the
GUT scale ($\sim 10^{16}$~GeV). Since the ${\cal U}(1)$ symmetry is
anomalous, the Fayet-Iliopoulos term 

\beq
\xi \int d^4\theta V_A
\label{fi}
\eeq
will be
generated \cite{fi}, where, in string theory
\cite{xi}

\begin{equation}
\xi =\frac{g_A^2M_P^2}{192\pi^2}{\rm Tr}Q~.
\label{xi}
\end{equation}
The $D_A$-term will have the form
\begin{equation}
\frac{g_A^2}{8}D_A^2=\frac{g_A^2}{8}
\left(\Sigma Q_i|\phi_i |^2+\xi \right)^2~,
\label{da}
\end{equation}
where $Q_i$ is the `anomalous' charge of $\phi_i $ superfield.
With opposite signs of $\xi $ and $Q_X$, the cancellation of (\ref{da})
fixes a non-zero VEV for the scalar component of $X$,

\beq
\langle X\rangle =\left(-\frac{\xi }{Q_X} \right)^{1/2}~.
\label{xvev}
\eeq

We will assume that the scale of ${\cal U}(1)$ symmetry breaking
is  

\beq
\frac{\langle X\rangle }{M_P}\equiv \epsilon \simeq 0.22~.
\label{eps}
\eeq
In refs. \cite{gia} the anomalous ${\cal U}(1)$ symmetry was considered as
a
mediator of SUSY breaking, while in refs. \cite{anu1} the anomalous
Abelian
symmetries were exploited as flavor symmetries for a natural understanding
the hierarchies of fermion masses and mixings.
The parameter $\epsilon $ 
is an important expansion parameter in our scheme. Below we will express the magnitudes of
Yukawa couplings and CKM matrix elements in terms of $\epsilon $ and
$\epsilon_G$ (see (\ref{vevz})).

\subsection{$\mu $-Term, Charged Fermion Masses and Mixings\\
and Related Issues}

We start our considerations with the lepton sector and 
introduce an additional
three families of vector-like supermultiplets $\overline E+E$. These
states, together with $e^c$, $l$, $h_d$, $h_u$, have flavor-universal
${\cal R}$ charges:

$$
R_{e^c}=\frac{6R}{5}~,~~~R_{l}=-\frac{9R}{10}~,~~~
R_{\bar E}=\frac{7R}{10}~,~~~R_{E}=\frac{3R}{10}~,~~~
$$
\beq
R_{h_d}=-\frac{R}{2}~,~~~~~~~~R_{h_u}=\frac{17R}{10}~,~~~
\label{rch}
\eeq
while the ${\cal U}(1)$ assignment has flavor dependent structure:

$$
Q_{e^c_1}=2q+4Q_X~,~~~Q_{e^c_2}=Q_{e^c_3}=2q+\frac{11}{2}Q_X~,~~~
Q_{l_1}=-3q-3Q_X~,~~~
$$
$$
Q_{l_2}=-3q-\frac{3}{2}Q_X~,~~~Q_{l_3}=-3q+\frac{1}{2}Q_X~,~~~
Q_{E_1}=-6Q_X~,
$$
$$
Q_{E_2}=Q_{E_3}=-\frac{15}{2}Q_X~,~~~Q_{\overline E_1}=-2Q_X~,~~~
Q_{\overline E_2}=\frac{3}{2}Q_X~,
$$
\beq
Q_{\overline E_3}=-\frac{1}{2}Q_X~,~~~Q_{h_d}=-2q~,~~~
Q_{h_u}=6q+6Q_X~.
\label{u1ch}
\eeq
With the prescriptions (\ref{rch}), (\ref{u1ch}), and taking into account 
(\ref{cond}), one observes that the direct coupling $h_dh_u$ is forbidden
to all orders, and the tree level Yukawa couplings $e^clh_d$ also vanish.
The presence of $\overline E+E$ states is therefore crucial. 

As we see from (\ref{u1ch}), the states of the second and third generations
have non-integer $Q_X$ charges and therefore will not participate in
the type of couplings in (\ref{bilin}), 
while the states from the first family will be relevant for
the generation of the  $\mu $-term.
The relevant couplings have the following matrix representation:

\begin{equation}
\begin{array}{ccc}
 & {\begin{array}{ccc}
\hspace{-5mm}~~l_1~~~~~~~~~~ & \,\,~~~~h_d  & \,\,~~~~~~~~~~~E_1

\end{array}}\\ \vspace{2mm}
\begin{array}{c}
e^c_1 \\ h_u \\ \bar E_1
 \end{array}\!\!\!\!\! &{\left(\begin{array}{ccc}
\,\,0~~  &\,\, \left(\frac{X}{M_P}\right)^2E &
\,\,\left(\frac{X}{M_P}\right)^2 h_d
\\
\,\,0~~   &\,\,0~~  &
\,\,Z\left(\frac{\overline Z}{M_P}\right)^7~~
 \\
\,\,\frac{Z^3}{M_P^2}\left(\frac{X}{M_P}\right)^5 &\,\,
M_P\left(\frac{ZX}{M_P^2}\right)^2 
&\,\,M_P\left(\frac{X}{M_P}\right)^8~~
\end{array}\right) }~.
\end{array}  \!\!  ~~~~~
\label{modlep3}
\end{equation}
Substituting in (\ref{modlep3}) the VEVs of appropriate superfields
(\ref{vevz}), (\ref{eps}) and comparing (\ref{modlep3}) with (\ref{lep3}), 
from the expressions (\ref{lepmass}), (\ref{bilin1}), (\ref{mu1}) we
obtain

$$
h_u(\mu_0h_d+\mu_1 l_1)~,~~~
\mu \simeq M_P\frac{\epsilon_G^{10}}{\epsilon^6}\simeq 200~{\rm
GeV}~,~~~~~
$$
\beq
\mu_1 \sim \epsilon_G\epsilon^3\mu ~,~~
\label{modmu}
\eeq

\beq
\lambda_e \sim \frac{\epsilon_G^3}{\epsilon }~.
\label{modmas1}
\eeq

We see that the $\mu $-term has just the desired magnitude. Furthermore, from
(\ref{modmas1}) one can verify that the MSSM parameter $\tan \beta $
is close to unity.
The bilinear $R$-parity violating $\mu_1$-term in (\ref{modmu})
will cause the LSP to be unstable \cite{ber}, so that an alternative
candidate for cold dark matter must be found.

Next we consider the couplings which are relevant for the two heavier
generations:

\begin{equation}
\begin{array}{cccc}
 & {\begin{array}{cccc}
\hspace{-5mm}~~l_2~~~~~~~~ & \,\,~~~~~l_3~~~  & 
\,\,~~~~~~~~E_2~~ &\,\,~~~~~~~~~~E_3
\end{array}}\\ \vspace{2mm}
\begin{array}{c}
e^c_2 \\ e^c_3 \\ \bar E_2 \\ \bar E_3
 \end{array}\!\!\!\!\! &{\left(\begin{array}{cccc}
\,\,0~~  &\,\, 0 &\,\,\left(\frac{X}{M_P}\right)^2 h_d &\,\,
\left(\frac{X}{M_P}\right)^2 h_d
\\
\,\,0~~   &\,\,0  &
\,\,\left(\frac{X}{M_P}\right)^2 h_d &\,\,
\left(\frac{X}{M_P}\right)^2 h_d
\\
\,\,M_P\left(\frac{Z}{M_P}\right)^3 &\,\,0 &
\,\,M_P\left(\frac{X}{M_P}\right)^6~~ &\,\,
M_P\left(\frac{X}{M_P}\right)^6
\\
\,\,\frac{X^2}{M_P}\left(\frac{Z}{M_P}\right)^3 &\,\,
M_P\left(\frac{Z}{M_P}\right)^3 &\,\,
M_P\left(\frac{X}{M_P}\right)^8 &\,\,
M_P\left(\frac{X}{M_P}\right)^8
\end{array}\right) }~.
\end{array}  \!\!  ~~~~~
\label{lep0}
\end{equation}
Integration of the heavy $(\overline E+E)_{2,3}$ states yields

\begin{equation}
\begin{array}{cc}
 & {\begin{array}{cc}
\hspace{-5mm}l_2~ & \,\,l_3~~~

\end{array}}\\ \vspace{2mm} 
\begin{array}{c}
e^c_2\\ e^c_3
 \end{array}\!\!\!\!\! &{\left(\begin{array}{cc}
\,\,\epsilon^2  &\,\,1
\\ 
\,\,\epsilon^2  &\,\,1~~
\end{array}\right)\frac{\epsilon_G^3}{\epsilon^6}h_d }~.
\end{array}  \!\!  ~~~~~
\label{lep1} 
\end{equation}
From (\ref{lep1}), (\ref{modmas1}) we find
$$
\lambda_{\tau }\sim ~\frac{\epsilon_G^3}{\epsilon^6}\sim 10^{-2},~~~~~~~
\tan \beta \sim 1~,
$$
\beq
\lambda_e:\lambda_{\mu }:\lambda_{\tau }\sim
\epsilon^5:\epsilon^2:1~,
\label{lep2}
\eeq
which is indeed the desirable hierarchical structure for the Yukawa couplings of
the charged leptons.

Turning to the quark sector, for generating the down quark masses we
introduce three pairs of $\overline D^c+D^c$. With the
transformation properties:

\beq
R_{q}=R_{e^c}~,~~R_{d^c}=R_l~,~~R_{D^c}=R_E~,~~
R_{\overline D^c}=R_{\overline E}
\label{rchq1}
\eeq

$$
Q_{q_1}=2q+\frac{5}{2}Q_X~,~~Q_{q_2}=2q+\frac{7}{2}Q_X~,~~
Q_{q_3}=2q+\frac{11}{2}Q_X~,~~
$$
$$
Q_{d^c_1}=-3q-\frac{1}{2}Q_X~,~~
Q_{d^c_2}=Q_{d^c_3}=-3q+\frac{1}{2}Q_X~,~~
$$
\beq
Q_{D^c_i}=-\frac{15}{2}Q_X~,~~~~~
Q_{\overline D^c_i}=-\frac{1}{2}Q_X~,
\label{u1chq1}
\eeq
the mass matrix relevant for down quark masses has the form

\begin{equation}
\begin{array}{cc}
 & {\begin{array}{cc}
\hspace{-5mm}~~d^c~ & \,\,~~~~D^c

\end{array}}\\ \vspace{1mm} 
\begin{array}{c}
q\\ 
\overline D^c
 \end{array}\!\!\!\!\! &{\left(\begin{array}{cc}
\,\,0  &\,\,\hat{A} h_d
\\ 
\,\,\hat{M}_{\overline D^cd^c}~~   &\,\,\hat{M}_{D^c}~~
\end{array}\right) }~,
\end{array}  \!\!  ~~~~~
\label{down} 
\end{equation}
where

\begin{equation}
\begin{array}{ccc}
\hat{A}=~~ \\
\end{array}
\hspace{-6mm}\left(
\begin{array}{ccc}
\epsilon^3& \epsilon^3 & \epsilon^3 \\
\epsilon^2 & \epsilon^2& \epsilon^2 \\
1& 1 & 1 \end{array}
\right)\epsilon^2 ~,~~
\begin{array}{ccc}
\hat{M}_{\overline D^cd^c}=~~ \\
\end{array}
\hspace{-6mm}\left(
\begin{array}{ccc}
\epsilon & 1 & 1 \\
\epsilon & 1 & 1 \\
\epsilon & 1 & 1 \end{array}
\right)\frac{Z^3}{M_P} ~,
\label{mats1}
\end{equation}

\begin{equation}
\hat{M}_{D^c}^{ij}=M_P\left(\frac{X}{M_p}\right)^8\alpha^{ij}~
\label{gd}
\end{equation}
($\alpha^{ij}$ are dimensionless couplings of order unity).
Integrating out the heavy $\overline D^c+D^c$ states gives

\begin{equation}
\begin{array}{ccc}
 & {\begin{array}{ccc}
~d^{c}_1& \,\,~~d^{c}_2  & \,\,~d^{c}_3~~~~~~~
\end{array}}\\ \vspace{2mm}
\hat{m}_d=A\hat{M}_{D^c}^{-1}\hat{M}_{\overline D^cd^c}h_d\simeq
\begin{array}{c}
q_1\\ q_2 \\q_3
 \end{array}\!\!\!\!\! &{\left(\begin{array}{ccc}
\,\,\epsilon^4  &\,\,~~\epsilon^3 &
\,\,~~\epsilon^3
\\
\,\,\epsilon^3   &\,\,~~\epsilon^2  &
\,\,~~\epsilon^2
 \\
\,\, \epsilon &\,\,~~ 1  &\,\,~~1
\end{array}\right) \frac{\epsilon_G^3}{\epsilon^6} h_d}~,
\end{array}  \!\!  ~~~~~
\label{down2}
\end{equation}
which upon diagonalization yields

\beq
\lambda_b\sim ~\frac{\epsilon_G^3}{\epsilon^6}\sim 10^{-2},~~~
\lambda_d:\lambda_s:\lambda_b\sim
\epsilon^4:\epsilon^2:1~,
\label{down0}
\eeq
the desired hierarchies between the three families of down quarks.

Before discussing the up quark sector, let us note that the heavy
decoupled states ($\overline E+E$ doublets and $\overline D^c+D^c$
triplets) have masses

$$
m_{d_1}\simeq m_{d_3}\simeq M_P\epsilon^8~,~~~~~
m_{d_2}\simeq M_P\epsilon^6~,
$$
\beq
m_{t_1}\simeq m_{t_2}\simeq m_{t_3}\simeq M_P\epsilon^8~ ,\label{massdt}
\eeq
and constitute three $\bar 5+5$ states of $SU(5)$.
This allows the possibility to retain the successful unification of
the three gauge
couplings, and also obtain an improved value for $\alpha_s(M_Z)$

\beq
\alpha_s^{-1}=\left(\alpha_s^0\right)^{-1}-\frac{9}{14\pi }
\ln \frac{m_{t_1}m_{t_2}m_{t_3}}{m_{d_1}m_{d_2}m_{d_3}}~,
\label{alphas}
\eeq
where $\alpha_s^0$ is the strong coupling, calculated at $M_z$ in minimal SUSY
$SU(5)$. Taking $\left(\alpha_s^0\right)^{-1}=1/0.126$ \cite{lan}
and using (\ref{massdt}), from (\ref{alphas}) we obtain
$\alpha_s=0.117$, which is in excellent agreement with world average value
\cite{pd}. 

Turning to the up quark sector, we prescribe the following transformation
properties to the $u^c$
states

$$
R_{u^c_i}=-\frac{19}{10}R~,~~~~~Q_{u^c_1}=-8q-\frac{29}{2}Q_X~,~~
$$
\beq
Q_{u^c_2}=-8q-\frac{25}{2}Q_X~,~~~Q_{u^c_2}=-8q-\frac{23}{2}Q_X~.~~
\label{chu}
\eeq
The up quark mass matrix will have the structure:

\begin{equation}
\begin{array}{ccc}
 & {\begin{array}{ccc}
~~u^{c}_1& \,\,~~u^{c}_2  & \,\,~u^{c}_3~~~~~~
\end{array}}\\ \vspace{2mm}
\hat{m}_u\simeq
\begin{array}{c}
q_1\\ q_2 \\q_3
 \end{array}\!\!\!\!\! &{\left(\begin{array}{ccc}
\,\,\epsilon^6  &\,\,~~\epsilon^4 &
\,\,~~\epsilon^3
\\
\,\,\epsilon^5  &\,\,~~\epsilon^3  &
\,\,~~\epsilon^2
 \\
\,\, \epsilon^3 &\,\,~~ \epsilon  &\,\,~~1
\end{array}\right)h_u}~,
\end{array}  \!\!  ~~~~~
\label{up}
\end{equation}
whose diagonalization yields

\beq
\lambda_t\sim 1,~~~
\lambda_u:\lambda_c:\lambda_t\sim
\epsilon^6:\epsilon^3:1~.
\label{upmass}
\eeq
From (\ref{down2}) and (\ref{up}) one can also estimate the CKM
matrix
elements

\beq
V_{us}\sim \epsilon~,~~~~V_{cb}\sim \epsilon^2~,~~~~
V_{ub}\sim \epsilon^3~.
\label{ckm}
\eeq
which is in very good agreement with observations.

It is worth noting that since the states $d^c$, $u^c$ have non-integer $Q_X$
charges, the baryon number violating trilinear 
couplings $u^cd^cd^c$ are forbidden. This is a consequence of the fact that in quark
sector, due to ${\cal U}(1)$ symmetry, $R$-parity emerges
automatically.

The Planck scale suppressed baryon number violating
$d=5$ operators

$$
\frac{1}{M_P}~qqql~,~~~~\frac{1}{M_P}~qqqE~,~~~~\frac{1}{M_P}~qqqh_d~,
$$
\beq
\frac{1}{M_P}~u^cu^cd^ce^c~,~~~~~~~~~\frac{1}{M_P}~u^cu^cD^ce^c~,~~
\label{d5ops}
\eeq  
are also forbidden by the ${\cal R}$ symmetry
\cite{d5ops}.
Thanks to the ${\cal R}\times {\cal U}(1)$ symmetry, baryon number
is conserved in our scheme \cite{ben}
\footnote{Different mechanism for baryon number
conservation in $R$-parity violating GUT scenario was considered in 
\cite{joni}, where a missing VEV solution of the GUT adjoint was
applied.}.

As far as lepton number violation is concerned, the couplings
$e^cll$, $e^cEE$ and $qD^cE$ are also forbidden by
${\cal R}\times {\cal U}(1)$. The $qd^cl$ type operator has
at least the suppression factor $\left(\frac{Z}{M_P}\right)^4$, which satisfies
all phenomenological bounds \cite{pb} and will not have any significant
contributions to neutrino masses (that emerge radiatively through this
coupling).

Some other allowed operators are

\beq
\frac{Z}{M_P}\Gamma_{ijk}~ e^c_iE_jl_k~,~~
\frac{Z}{M_P}\Gamma_{ijk}'~ q_id^c_jE_k~,~~
\frac{Z}{M_P} \Gamma_{ijk}''~q_iD^c_jl_k~~~
\label{lnum}
\eeq
($\Gamma $, $\Gamma' $ and $\Gamma'' $ are family dependent couplings),
and they all involve the decoupled heavy states.
They give rise to the lepton number rotating operators  

\beq
\lambda_{ijk}~e^cl_jl_k~,~~~~~\lambda_{ijk}'~q_id^c_jl_k~.
\label{treelin}
\eeq 
Taking into account 
(\ref{modlep3}), (\ref{lep0}), (\ref{down})-(\ref{gd}) one can verify
that

$$
E_3\supset
\frac{\epsilon_G^3}{\epsilon^8}\left(\epsilon^2l_2+l_3\right)~,~~
E_2\supset \frac{\epsilon_G^3}{\epsilon^6}l_2~,~~
E_1\supset \frac{\epsilon_G^3}{\epsilon^3}l_1~,~~
$$
\beq
D^c_i\supset \frac{\epsilon_G^3}{\epsilon^8}
\left(\epsilon d^c_1+d^c_{2,3}\right)~.
\label{weights0}
\eeq
The $\Gamma$ factors in (\ref{lnum}) are expressed through appropriate powers
of $X$, depending on the appropriate superfield and can be selected
from the prescriptions (\ref{u1ch}), (\ref{u1chq1}). From this, taking
into
account (\ref{lnum}), (\ref{weights0}), 
the non-zero  $\lambda $ and $\lambda'$ suppression factors are
(let us note that $\lambda_{ijk}=-\lambda_{ikj}$)

$$
\lambda_{123}\sim \lambda_{213}\sim \lambda_{313}\sim
\lambda_{321}'\sim \lambda_{331}'\sim \frac{\epsilon_G^4}{\epsilon^3 }~, 
$$
$$
\lambda_{212}\sim \lambda_{312}\sim \lambda_{221}'\sim
\lambda_{231}'\sim \frac{\epsilon_G^4}{\epsilon }~,
$$
\beq
\lambda_{111}'\sim \epsilon_G^4\epsilon~,~~
\lambda_{121}'\sim \lambda_{131}'\sim \lambda_{211}'\sim
\epsilon_G^4~,~\lambda_{311}'\sim \frac{\epsilon_G^4}{\epsilon^2}~.
\label{lambdas}
\eeq 

All other trilinear lepton number violating couplings which are
not presented in (\ref{lambdas}) are zero due to 
${\cal R}\times {\cal U}(1)$ symmetry.
We note that $\lambda, \lambda'$ are suppressed at the required level, so 
that the phenomenological bounds \cite{pb} obtained from different processes 
are satisfied. The contributions (through radiative corrections)
from these couplings to the neutrino masses are also negligible.

The $R$-parity violating bilinear and trilinear couplings make the LSP
unstable.
In our model the dominant contribution to LSP decay comes
from the bilinear 
$\mu_1$-term in (\ref{modmu}). The lifetime for decay into three fermions
is given by

\beq
\tau_{\chi }^{-1}=\mu_1^2Z_{\chi \tilde{h}}^2
\left(\frac{1}{4}+\sin^2 \theta_W+\frac{4}{3}\sin^4 \theta_W \right)
\frac{G_F^2m_{\chi }^3}{192\pi^3 }~,
\label{time}
\eeq
and for $m_{\chi }\sim 100$~GeV, taking into account (\ref{modmu}), 
we obtain $\tau_{\chi }\sim 10^{-19}$~sec. 
Therefore, the LSP is cosmologically irrelevant.

\subsection{Neutrino Oscillations and Leptogenesis}

In this section we investigate the neutrino sector of our model 
and attempt to
accommodate the recent atmospheric \cite{atm} and solar \cite{sol} neutrino
data.
Starting with atmospheric neutrinos, let us note that the 
prescription (\ref{u1ch})
of ${\cal U}(1)$ charges for $l_2, l_3$ permit us to realize
maximal mixing between $\nu_{\mu }$ and $\nu_{\tau }$ through the
mechanism described in \cite{maxmix}. Introducing two right handed 
${\cal N}_{2,3}$ states with transformation properties

\beq
R_{{\cal N}_{2, 3}}=\frac{R}{5}~,~~~~
Q_{{\cal N}_2}=-3q-\frac{19}{2}Q_X~,~~~~
Q_{{\cal N}_3}=-3q-\frac{9}{2}Q_X~,
\label{nch}
\eeq
with the condition
\beq
Q_X=-\frac{5}{14}q~,
\label{cond}
\eeq
the relevant couplings have the desirable textures:

\begin{equation}
\begin{array}{cc}
 & {\begin{array}{cc}
~~{\cal N}_2&\,\,{\cal N}_3~~~
\end{array}}\\ \vspace{2mm}
\begin{array}{c}
l_2\\l_3\\ 

\end{array}\!\!\!\!\! &{\left(\begin{array}{cc}
\, \epsilon^5~ &
\,\,~1
\\
\, \epsilon^3 &
\,\,~0
\end{array}\right)h_u }
\end{array}  \!\!~,~~~~~~~~~~~
\begin{array}{cc}
 & {\begin{array}{cc}
~{\cal N}_2&\,\,
{\cal N}_3~~~~
\end{array}}\\ \vspace{2mm}
\begin{array}{c}
{\cal N}_2 \\ {\cal N}_3

\end{array}\!\!\!\!\! &{\left(\begin{array}{ccc}
\, \epsilon^5~~
 &\,\,1
\\
\, 1~~
&\,\,0
\end{array}\right)\frac{Z^2\overline Z}{M_P}}
\end{array} ~.~~
\label{lN}
\end{equation}
After integrating out the heavy ${\cal N}_{2,3}$ states, the neutrino mass matrix
for $\nu_{\mu }, \nu_{\tau }$ will have the quasi-degenerate form

\begin{equation}
\begin{array}{cc}
 & {\begin{array}{cc}
\hspace{-5mm}~~\nu_{\mu } & \,\,\nu_{\tau }~

\end{array}}\\ \vspace{2mm} 
\begin{array}{c}
\nu_{\mu }\\ \nu_{\tau }
 \end{array}\!\!\!\!\! &{\left(\begin{array}{cc}
\,\,\epsilon^2 &\,\,1
\\ 
\,\,1~~   &0
\end{array}\right)m }~,
\end{array}  \!\!  ~~~~~
m=\frac{\epsilon^3h_u^2}{M_P\epsilon_G^3}~,
\label{lep4} 
\end{equation}
with

\beq
m_{\nu_2}\simeq m_{\nu_3}\simeq m\simeq 0.13~{\rm eV}~.
\label{deg}
\eeq
For the atmospheric neutrino oscillation
parameters we find
$$
\Delta m_{23}^2=2m^2\epsilon^2 \simeq 2\cdot 10^{-3}~{\rm eV}^2~,
$$
\begin{equation}
\sin^2 2\theta_{\mu \tau } =1-{\cal O}(\epsilon^4)~.
\label{atm}
\end{equation}

The resolution of the solar neutrino puzzle in our scenario requires the introduction
of a light sterile neutrino state. We have arranged the lepton
sector in such a way as to avoid mixing of the first generation with the second
and third generations. This was necessary because $R$-parity
violation (through bilinear terms) in the sector of heavy generations
could create unacceptably heavy neutrinos. We violated $R$-parity
in the sector of first generation and generated the terms $\mu $, $\mu_1$
in (\ref{modmu}). Due to ${\cal R}\times {\cal U}(1)$  there arises
the following hierarchy between $\mu $ and $\mu_1$, and therefore between
$B_{\mu }$ and $B_{\mu_1}$ as well,

\beq
\frac{\mu_1}{\mu }=\frac{B_{\mu_1}}{B_{\mu }}\sim 
\epsilon_G\epsilon^3~.
\label{hier0}
\eeq

The alignment holds at $M_P$ since we are working within the
framework of minimal $N=1$ SUGRA. This alignment is violated due
to renormalization between $M_P$ and mass scales of `$E, D^c$' states and,
taking into account (\ref{mis1}), (\ref{modlep3}),
(\ref{lep0}), (\ref{mats1}), we expect

\beq
\frac{\Delta \mu }{\mu }\sim \frac{6\epsilon^2}{16\pi^2}\ln \epsilon^8~.
\label{renorm}
\eeq
The misalignment parameter is estimated to be 

\beq
\sin \xi \sim \frac{\mu_1}{\mu}\frac{6\epsilon^4}{16\pi^2}\ln \epsilon^8
\sim \epsilon_G\epsilon^3~\frac{6\epsilon^4}{16\pi^2}\ln \epsilon^8
\simeq 10^{-7}~,
\label{mismodel}
\eeq
and the mass of the `$\nu_e$' state (see (\ref{massnu})) is given by

\beq
m_{\nu_e}\simeq 10^{-3}~{\rm eV}.
\label{nuemass}
\eeq

As we mentioned above, the state $\nu_e$ does not mix with $\nu_{\mu, \tau}$
(which in any case are too heavy for the solar neutrino puzzle).
We introduce a sterile state 
$\nu_s$ with the transformation properties:

\beq
R_{\nu_s }=\frac{R}{5}~,~~~~~~~Q_{\nu_s }=-3q-26Q_X~.
\label{stch}
\eeq
Taking into account (\ref{cond}), the relevant couplings are
\beq
W_{\nu_s }=\left(\frac{X}{M_P}\right)^{23}l_1\nu_sh_u+
\frac{Z^2\overline Z}{M_P^2}\left(\frac{X}{M_P}\right)^{38}\nu_s^2~,
\label{supst}
\eeq
from which we have

\beq
m_{\nu_s}=M_P\epsilon_G^3\epsilon^{38}\simeq 2.5\cdot 10^{-4}~{\rm
eV},~~~~~
m_{\nu_s \nu_e}=\epsilon^{23}h_u\simeq 10^{-4}~{\rm eV}~.
\label{masst}
\eeq
Note that $\nu_s$ is kept light (in (\ref{supst}), (\ref{masst})) 
by the symmetry ${\cal R}\times {\cal U}(1)$ \cite{st, d5ops}.

Taking into account (\ref{nuemass}), (\ref{masst}), for the solar neutrino
oscillation
parameters we will have

$$
\Delta m_{\nu_e \nu_s}^2\simeq m_{\nu_e}^2
\sim  10^{-6}~{\rm eV}^2~,
$$
\begin{equation}
\sin^2 2\theta_{es}\simeq
4\left( \frac{m_{\nu_s\nu_e}}{m_{\nu_e }}\right)^2
\sim 10^{-2} ~,
\label{sol}
\end{equation}
which explains the solar $\nu_e$ deficit through the small angle MSW 
oscillations. The
tree level induced $\nu_e$ mass (\ref{nuemass}), whose origin 
lies in the $R$-parity
violating bilinear (\ref{modmu}) term, plays a crucial role in (\ref{sol}).

Before concluding, let us note that even though  baryon number is
perturbatively conserved in
our model, the observed baryon asymmetry can be explained
by first creating lepton asymmetry \cite{lepas} through
the out of equilibrium decay of the right handed
neutrinos ${\cal N}_{2,3}$. The electroweak sphalerons \cite{spa}
would partially transform the lepton asymmetry to the observed baryon 
asymmetry.
The out-of-equilibrium
condition reads

\beq
\Gamma \stackrel{<}{_\sim }KH=1.7K\sqrt{g_*}\frac{T^2}{M_P}~,
\label{equil}
\eeq 
where $\Gamma $ is the decay rate of ${\cal N}_{1,2}$ states into  
leptons, $H$ is Hubble's constant, $g_*$ is the effective number of
massless degree of freedom, and $K=1-10^{3}$. 
The heavy neutrino decay rate is

\beq
\Gamma_{{\cal N}_i} =\frac{(h^{+}h)_{ii}}{8\pi }M_{{\cal N}_i}~,
\label{width}
\eeq
where $h$ can be extracted from $hl{\cal N}h_u$ type couplings of 
(\ref{lN}). Using (\ref{width}) and taking into account (\ref{lN})
one can see that (\ref{equil}) is satisfied for $K=10^3$.

In addition to the out-of-equilibrium condition (\ref{equil}) we need CP
violation,
which is necessary for generating the asymmetry. 
In our model	
the states ${\cal N}_2$ and ${\cal N}_3$ are nearly
degenerate in mass. This, as shown in refs. \cite{res}, can be a very
convenient fact because of the resonance
enhancement of the CP asymmetry that it creates. 
On the other hand, the baryon-to-entropy density ratio $Y_B=n_B/s$,
which is created from lepton-to-entropy density ratio through sphalerons,
equals \cite{ratio}

\beq
Y_B\approx -\frac{1}{3}Y_L
\approx -\frac{1}{2K}\frac{\delta_{CP}}{g^*}~.
\label{ratbar}
\eeq
According to (\ref{ratbar}), for a large CP asymmetry $\delta_{CP}$
in order to have $Y_B\approx 10^{-10}$ one can take larger values for
$K$, so that (\ref{equil}) will be more readily satisfied.
Therefore, we conclude that all
of the conditions \cite{sak} for obtaining the baryon asymmetry of the
Universe
are satisfied within the framework of our scenario.

\end{document}